\documentclass{pasj01}
\Received{$\langle$2022 April 20$\rangle$}
\Accepted{$\langle$2022 August 22$\rangle$}
\Published{$\langle$publication date$\rangle$}

\usepackage{url}
\usepackage{lineno}
\usepackage{multirow}
\makeatletter
\def\@fnsymbol#1{\ensuremath{\ifcase#1\or \dagger\or \ddagger\or
   \mathsection\or \mathparagraph\or \|\or **\or \dagger\dagger
   \or \ddagger\ddagger \else\@ctrerr\fi}}
    \makeatother

\begin{document}

\title{Optical IFU Observations of GOALS Sample with KOOLS-IFU on Seimei Telescope: Initial results of 9 U/LIRGs at \boldmath$z <$ 0.04}

 \author{
   Yoshiki		\textsc{Toba}			\altaffilmark{1,2,3,4,5,$^\ast$\thanks{NAOJ Fellow}},
   Satoshi 		\textsc{Yamada}			\altaffilmark{6,2},
   Kazuya 		\textsc{Matsubayashi}	\altaffilmark{7,8}, 
   Koki 		\textsc{Terao}			\altaffilmark{9,10},    
   Aoi			\textsc{Moriya}			\altaffilmark{3},   
   Yoshihiro 	\textsc{Ueda}	   		\altaffilmark{2},   
   Kouji 		\textsc{Ohta}			\altaffilmark{2},
   Aoi			\textsc{Hashiguchi}		\altaffilmark{3},    
   Kazuharu	G.	\textsc{Himoto}			\altaffilmark{11}, 
   Hideyuki 	\textsc{Izumiura}		\altaffilmark{12}, 
   Kazuma 		\textsc{Joh}			\altaffilmark{11},
   Nanako		\textsc{Kato}			\altaffilmark{11}, 
   Shuhei		\textsc{Koyama}			\altaffilmark{7,5}, 
   Hiroyuki     \textsc{Maehara}		\altaffilmark{12}, 
   Rana			\textsc{Misato}			\altaffilmark{3},
   Akatoki 		\textsc{Noboriguchi}	\altaffilmark{13,10,11},
   Shoji		\textsc{Ogawa}			\altaffilmark{2},
   Naomi		\textsc{Ota}			\altaffilmark{3},
   Mio			\textsc{Shibata}		\altaffilmark{3},   
   Nozomu		\textsc{Tamada}			\altaffilmark{11}, 
   Anri			\textsc{Yanagawa}		\altaffilmark{3},  
   Naoki		\textsc{Yonekura}		\altaffilmark{11}, 
   Tohru		\textsc{Nagao}			\altaffilmark{5},
   Masayuki		\textsc{Akiyama}		\altaffilmark{10},    
   Masaru		\textsc{Kajisawa}		\altaffilmark{5,11},   
   Yoshiki		\textsc{Matsuoka}		\altaffilmark{5}
}
\email{yoshiki.toba@nao.ac.jp}

\altaffiltext{1}{National Astronomical Observatory of Japan, 2-21-1 Osawa, Mitaka, Tokyo 181-8588, Japan}
\altaffiltext{2}{Department of Astronomy, Kyoto University, Kitashirakawa-Oiwake-cho, Sakyo-ku, Kyoto 606-8502, Japan}
\altaffiltext{3}{Department of Physics, Nara Women's University, Kitauoyanishi-machi, Nara, Nara 630-8506, Japan}
\altaffiltext{4}{Academia Sinica Institute of Astronomy and Astrophysics, 11F of Astronomy-Mathematics Building, AS/NTU, No.1, Section 4, Roosevelt Road, Taipei 10617, Taiwan}
\altaffiltext{5}{Research Center for Space and Cosmic Evolution, Ehime University, 2-5 Bunkyo-cho, Matsuyama, Ehime 790-8577, Japan}
\altaffiltext{6}{RIKEN Cluster for Pioneering Research, 2-1 Hirosawa, Wako, Saitama 351-0198, Japan}
\altaffiltext{7}{Institute of Astronomy, Graduate School of Science, The University of Tokyo, 2-21-1 Osawa, Mitaka,Tokyo 181-0015, Japan}
\altaffiltext{8}{Okayama Observatory, Kyoto University, Honjo 3037-5, Kamogata-cho, Asakuchi, Okayama 719-0232, Japan}
\altaffiltext{9}{Subaru Telescope, National Astronomical Observatory of Japan, 650 North A'ohoku Place, Hilo, HI 96720, USA}
\altaffiltext{10}{Astronomical Institute, Tohoku University, Aramaki, Aoba-ku, Sendai, 980-8578, Japan}
\altaffiltext{11}{Graduate School of Science and Engineering, Ehime University, Bunkyo-cho, Matsuyama 790-8577, Japan}
\altaffiltext{12}{Okayama Branch Office, Subaru Telescope, National Astronomical Observatory of Japan, NINS, Kamogata, Asakuchi, Okayama 719-0232, Japan}
\altaffiltext{13}{School of General Education, Shinshu University, 3-1-1 Asahi, Matsumoto, Nagano 390-8621, Japan}


\maketitle

\begin{abstract}
We present ionized gas properties of 9 local ultra/luminous infrared galaxies (U/LIRGs) at $z <$ 0.04 through IFU observations with KOOLS-IFU on Seimei Telescope. 
The observed targets are drawn from the Great Observatories All-sky LIRG Survey (GOALS), covering a wide range of merger stages.
We successfully detect emission lines such as H$\beta$, [O{\,\sc iii}]$\lambda$5007, H$\alpha$, [N{\,\sc ii}]$\lambda\lambda$6549,6583, and [S{\,\sc ii}]$\lambda\lambda$6717,6731 with a spectral resolution of $R$ = 1500--2000, which provides (i) spatially-resolved ($\sim$200--700 pc) moment map of ionized gas and (ii) diagnostics for active galactic nucleus (AGN) within the central $\sim$3--11 kpc in diameter for our sample.
We find that [O{\,\sc iii}] outflow that is expected to be driven by AGN tends to be stronger (i) towards the galactic center and (ii) as a sequence of merger stage.
In particular, the outflow strength in the late-stage (stage D) mergers is about 1.5 times stronger than that in the early-state (stage B) mergers, which indicates that galaxy mergers could induce AGN-driven outflow and play an important role in the co-evolution of galaxies and supermassive black holes.
\end{abstract}


\section{Introduction}
\label{Intro}

In the last two decades, galaxy mergers have been recognized as a key phenomenon for understanding galaxy evolution (e.g., \cite{Conselice} and reference therein). 
Star formation (SF) and/or active galactic nucleus (AGN) activity in galaxies are often triggered by galaxy mergers\footnote{We note that major mergers are not the only triggering mechanism of SF/AGN activity. Non-merger processes such as secular mechanisms may also trigger SF/AGN activity (see e.g., \cite{Schawinski,Draper,Sharma}, and references therein) (see also Section \ref{D_stage}).}, which enhances the inflow of material from galactic scales into the close environments of the nuclear region.
The galaxy merger also enhances infrared (IR) luminosity ($L_{\rm IR}$\footnote{$L_{\rm IR}$ is empirically defined as the luminosity integrated over a wavelength range of 8--1000 $\mu$m (e.g., \cite{SM,Chary}).}) (e.g., \cite{Sanders98,Veilleux02,Imanishi,Imanishi09,Toba_15,Toba_16,Toba_17a}).
These are observed as luminous IR galaxies (LIRGs) and ultraluminous IR galaxies (ULIRGs) with $L_{\rm IR}$ greater than 10$^{11}$ and 10$^{12}$ $L_{\odot}$, respectively \citep{SM}.
Recent works also suggested that the relative strength of SF and AGN activity in U/LIRGs may be varied in the course of galaxy mergers (e.g., \cite{Narayanan,Ricci,Blecha,Yamada,Yamada21}).
In particular, radiation from an AGN is expected to interact with the interstellar medium (ISM) and lead to the ejection or heating of the gas, which could be tightly associated with the growth of the galaxy and its supermassive black hole (SMBH) (e.g., \cite{Cazzoli,Toba_17b,Harrison,Chen,Finnerty,Jun,Fluetsch}).
That AGN-driven outflow is often observed in U/LIRGs: recent observations reported that dusty IR luminous AGN often shows a strong ionized gas outflow, and its strength seems to be correlated with $L_{\rm IR}$ \citep{Bischetti,Toba_17c,Chen,Jun}.
Therefore, U/LIRGs are a good laboratory to examine how AGN-driven outflow could influence the host galaxy and its SMBH.
However, those works are often limited to handling a whole galaxy as a system.
To address, from a spatially-resolved point of view, how the SF and AGN activity could be enhanced/quenched as a function of merger sequence,  Integral Field Unit (IFU) observation is a quite powerful tool that may trace the gas kinematics and energetics (e.g., \cite{Bae17,Shin,Pan}). 

In this work, we focus on the Great Observatories All-sky LIRG Survey (GOALS; \cite{Armus}) that provides a complete, flux-limited (i.e., flux density at 60 $\mu$m $>$ 5.24 Jy) U/LIRG sample in the local universe.
In addition to their multi-wavelength follow-up observations, \citet{Stierwalt} divided the GOALS U/LIRGs into sub-samples regarding the merger stage, based on a visual inspection of the IRAC/Spitzer 3.6 $\micron$ images and/or higher resolution images taken by e.g., Hubble Space Telescope (see also \cite{Haan}).
Therefore, the GOALS sample is an ideal laboratory to investigate the role of galaxy mergers in the co-evolution of galaxies and SMBHs. 

In this paper, we present the initial results of our follow-up campaign of GOALS sample with an optical IFU, the Kyoto Okayama Optical Low-dispersion Spectrograph with optical-fiber IFU (KOOLS-IFU; \cite{Yoshida,Matsubayashi}) on the Okayama 3.8 m, Seimei Telescope \citep{Kurita}. 
The structure of this paper is as follows. 
Section \ref{s_DS} describes the sample selection, observations, and data reduction of observed targets.
In Section \ref{s_RD}, we present the results of our optical IFU observations and discuss the dependence of ionized gas outflow on distance from the galaxy center and merger stage.
We summarize the results of this work in Section \ref{s_C}.
Throughout this paper, the adopted cosmology is a flat universe with $H_0$ = 70 km s$^{-1}$ Mpc$^{-1}$, $\Omega_{\rm M}$ = 0.28, and $\Omega_{\Lambda}$ = 0.72, which are the same as those adopted in a series of GOALS papers that are relevant to this work (e.g., \cite{Armus,Rich}).

\section{Data and analysis}
\label{s_DS}

\subsection{Target selection}
The observed targets were drawn from the GOALS sample, which is itself a subset of the IRAS Revised Bright Galaxy Sample \citep{Sanders}, providing a 60 $\micron$ flux-limited sample of U/LIRGs at $z < 0.088$.
The GOALS sample has been extensively examined by multi-wavelength observations from X-ray (e.g., \cite{Iwasawa,Ricci,Torres-Alba,Yamada20,Ricci21,Yamada21}), ultraviolet (UV) (e.g., \cite{Howell,Petty}), optical-NIR (e.g., \cite{Haan,Kim,Linden17,Jin,Larson}), mid-IR (MIR)--far-IR (FIR)  (e.g., \cite{Inami,Stierwalt,Stierwalt14,Chu,Diaz17,Inami18,Yamada,Armus20}) to radio (e.g., \cite{Privon,Barcos,Yamashita,Herrero,Linden,Condon}) (see also \cite{Casey,U} for SED analysis).
We employed the merger classifications provided by \citet{Stierwalt}, which have five categories from {\bf N} (no signs of merger activity or massive neighbors), to stage {\bf A} (galaxy pairs prior to a first encounter), {\bf B} (post-first encounter with galaxy disks that are still symmetric and intact but with signs of tidal tails), {\bf C} (showing amorphous disks, tidal tails, and other signs of merger activity), and {\bf D} (two nuclei in a common envelope), as a sequence of merger stage (see Section 2.5 in \cite{Stierwalt} for more detail) (see also e.g., Figure 1 in \cite{Ricci} for an example image of each merger stage).

\begin{figure}[t]
\begin{center}
\includegraphics[width=0.48\textwidth]{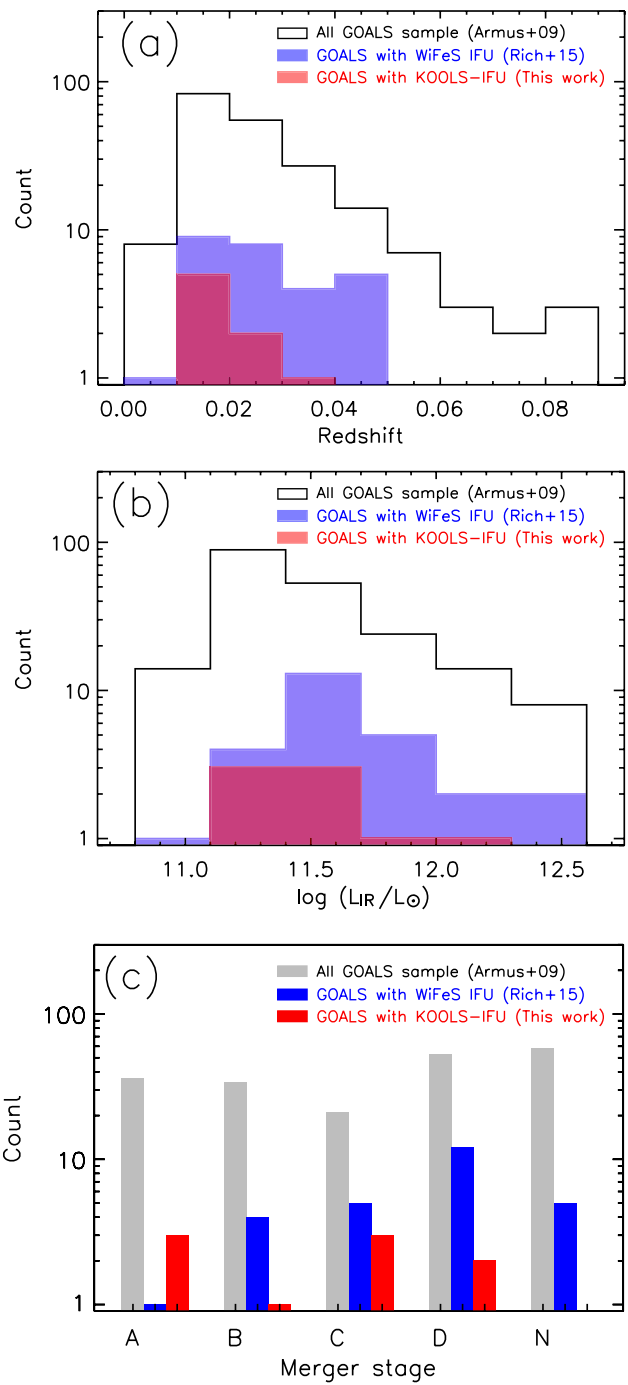}
\end{center}
\caption{Histograms of (a) redshift, (b) IR luminosity, and (c) merger stage of the GOALS sample (black/gray), GOALS sample with WiFeS provided in \citep{Rich} (blue), and GOALS sample with KOOLS-IFU, i.e., our sample (red).}
\label{hist}
\end{figure}

\begin{table*}[t]
\tbl{Basic properties of observed targets in GOALS sample.}{
\scalebox{0.85}[0.9]{
\begin{tabular}{llrrrrcccc}
\hline\hline
\multicolumn{1}{c}{Object Name}	&	\multicolumn{1}{c}{IRAS Name}	&	\multicolumn{1}{c}{R.A.}		&	\multicolumn{1}{c}{Decl.} 		&	\multicolumn{1}{c}{redshift} &	\multicolumn{1}{c}{$D_{\rm L}$}	& physical scale & 	$B$-mag  &	$\log\,(L_{\rm IR}/L_{\odot})$ 		&	Merger Stage	\\
			&				&	\multicolumn{1}{c}{(J2000.0)}	&	\multicolumn{1}{c}{(J2000.0)}	&		& (Mpc)		& (kpc/arcsec)	&		&			&					\\
\multicolumn{1}{c}{(1)}	& \multicolumn{1}{c}{(2)} &	\multicolumn{1}{c}{(3)}	& \multicolumn{1}{c}{(4)}	& \multicolumn{1}{c}{(5)}		& \multicolumn{1}{c}{(6)}	& \multicolumn{1}{c}{(7)}	&  \multicolumn{1}{c}{(8)}		&   \multicolumn{1}{c}{(9)}			&	 \multicolumn{1}{c}{(10)}				\\
\hline
  NGC 1614 & F04315$-$0840 & 04:34:00.03 & $-$08:34:44.57 & 0.01594 & 69.12 & 0.32 & 14.7 & 11.65 & D\\
  CGCG 468$-$002W & F05054+1718\_W & 05:08:19.71 & +17:21:48.09 & 0.01748 & 75.89 & 0.36 & 15.4 & (11.22) & B\\
  NGC 3690 West & F11257+5850\_W & 11:28:30.78 & +58:33:42.90 & 0.01017 & 43.90 & 0.21 & 11.8 & (11.93) & C\\
  NGC 3690 East & F11257+5850\_E & 11:28:33.39 & +58:33:46.40 & 0.01036 & 44.74 & 0.21 & 11.8 & (11.93) & C\\
  Mrk 273 & F13428+5608 & 13:44:42.07 & +55:53:13.17 & 0.03734 & 164.61 & 0.74 & 15.7 & 12.21 & D\\
  NGC 6786 & F19120+7320\_W & 19:10:53.75 & +73:24:36.60 & 0.02524 & 110.27 & 0.51 & 13.7 & (11.49) & C\\
  NGC 6921 & 20264+2533\_W & 20:28:28.84 & +25:43:24.19 & 0.01447 & 62.67 & 0.30 & 14.4 & (11.11) & A \\
  NGC 7674 & F23254+0830 & 23:27:56.70 & +08:46:44.24 & 0.02903 & 127.17 & 0.58 & 13.9 & 11.56 & A\\
  NGC 7679 & 23262+0314\_W & 23:28:46.67 & +03:30:40.99 & 0.01715 & 74.45 & 0.35 & 13.2 & 11.11 & A\\
\hline
\end{tabular}
}
}
\label{sample}
\begin{tabnote}
Columns: (1) object name from SIMBAD Astronomical Database or NASA/IPAC Extragalactic Database (NED). (2) IRAS name. (3--4) right ascension (R.A.) and declination (Decl.) (J2000.0). (5) spectroscopic redshift in NED. (6--7) luminosity distance and physical size in 1 arcsec in the assumed cosmology. (8) $B$-band magnitude from SIMBAD or NED. (9) IR luminosity from \citet{Armus}, where values in brackets denote the total IR luminosity for a system.  (10) merger stage from \citet{Stierwalt}.
\end{tabnote}
\end{table*}

\begin{figure*}
\begin{center}
\includegraphics[width=0.95\textwidth]{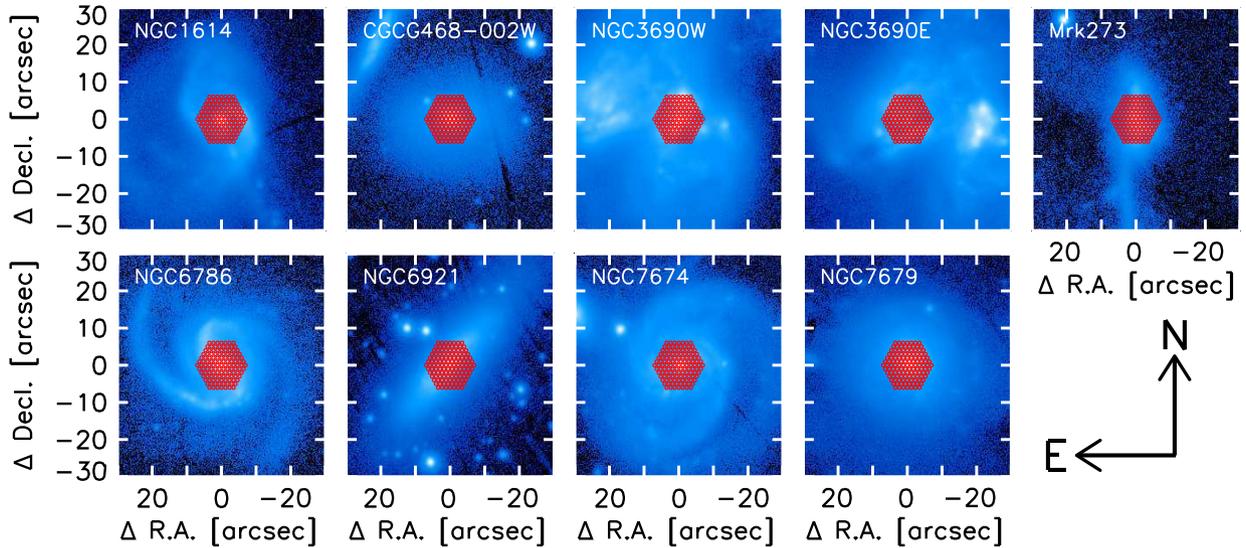}
\end{center}
\caption{The $i$-band postage stamps ($60\arcsec \times 60\arcsec$) of our sample taken by Pan-STARRS. The footprint of the KOOLS-IFU with 127 fibers is overlaid as red circles. R.A. and decl. are relative coordinates with respect to those in Table \ref{sample} in which ($\Delta$R.A, $\Delta$Decl.) = (0, 0) also corresponds to the  center of the field of view for KOOLS-IFU (i.e., fiber ID = 37, see Figure \ref{fiberID}). North is up and east is left in all images.}
\label{image}
\end{figure*}

We selected 9 objects as the target by taking into account the magnitude in optical bands ($r$-mag $<$ 16.0), merger stage, previous IFU observations, and visibility from the Seimei telescope.
The basic information of observed targets is summarized in Table \ref{sample}.
Figure \ref{hist} shows distributions of redshift, IR luminosity, and merger stage of our sample. 
We also present those distributions for all the GOALS sample \citep{Armus} and sub-sample reported in \citet{Rich}, who performed the IFU observations of 27 southern U/LIRGs from the GOALS sample with the Wide Field Spectrograph (WiFeS) installed on a 2.3 m telescope that capability is similar to KOOLS-IFU.
We confirm from Figure \ref{hist} that our sample is widely distributed in terms of IR luminosity and merger stage while that is biased toward low-$z$ ($z < 0.04$). 
Figure \ref{image} shows $i$-band images of our sample taken by Pan-STARRS\footnote{\url{https://ps1images.stsci.edu/cgi-bin/ps1cutouts}} \citep{Chambers}.

We note that the WiFeS-GOALS IFU survey \citep{Rich11,Rich12,Rich}, the VLT-VIMOS IFU survey \citep{Arribas}, the PMAS IFU survey \citep{Alonso}, the INTEGRAL IFU survey \citep{Garcia}, VLT-SINFONI IFU survey \citep{Piqueras}, Physics of ULIRGs with MUSE and ALMA (PUMA; \cite{Perna}), and  the Keck OSIRIS AO LIRG Analysis (KOALA; \cite{U19}) observed (a part of) GOALS sample (see also \cite{Garcia06,Rodriguez,Bellocchi,Arribas_14,Kakkad,Perna22}).
In particular, the WiFeS-GOALS IFU survey provides specially-resolved maps for emission-line ratios based on various emission lines, such as H$\beta$, [O{\,\sc iii}]$\lambda$5007, H$\alpha$, [N{\,\sc ii}]$\lambda\lambda$6549,6583, and [S{\,\sc ii}]$\lambda\lambda$6716,6731 that are also targeted lines in this work (see Section \ref{s_BPT}).
The sample in this survey is observable from the southern hemisphere, and thus our sample (that is observable from the northern hemisphere) is complementary to those reported in \citet{Rich}.
For the remaining projects, those basically focused on one emission line (e.g., H$\alpha$) and spatially-resolved emission-line ratios were not well investigated\footnote{We also confirmed that our targets have not yet been cataloged in the Calar Alto Legacy Integral Field Area Survey (CALIFA; \cite{Sanchez}) Data Release (DR) 3 \citep{Sanchez16}, the Sydney-AAO Multi-object Integral-field spectrograph (SAMI) galaxy survey \citep{Bryant} DR3 \citep{Croom}, and the Sloan Digital Sky Survey (SDSS: \cite{York}) Mapping Nearby Galaxies at Apache Point Observatory survey (MaNGA; \cite{Bundy}) DR17 \citep{Abdurrouf}.}.

\begin{table*}
\tbl{Observation log in 2019B, 2020A, and 2020B semester.}{
\scalebox{0.85}[0.9]{
\begin{tabular}{llrccl}
\hline\hline
\multicolumn{1}{c}{Name}	& \multicolumn{1}{c}{IRAS Name} &	 \multicolumn{1}{c}{Observing date} &		\multicolumn{2}{c}{Exposure time (minutes)}  & \multicolumn{1}{c}{Standard star} \\
			&			&					&		VPH 495	& VPH 683	&	\\	
\hline
CGCG 468-002W	&	F05054+1718\_W	    & Oct. 8, 2019  & 	40 	&	40  & EGGR247 (=G191--B2B)	\\
NGC 1614		&	F04315-0840         & Oct. 9, 27, 2019  & 50 &  50  & HR9087, HD74280 (=HR3454)  \\
NGC 7679		&	23262+0314\_W	    & Oct. 25, 2019 &	30	&	30  & HR7596 \\
\hline
NGC 3690 West &F11257+5850\_W	   	& Apr. 20,21, 2020 &	40	&	40	& HR5501, EGGR98 (=HZ43)  \\
NGC 3690 East &F11257+5850\_E     	& Apr. 22,23, 2020 &  	36  &	38 	& EGGR98 (=HZ43) \\
Mrk 273			&	F13428+5608		& Apr. 23, 2020 &   	36	&	36  & HR5501 \\
\hline
NGC 6786		&	F19120+7320\_W	    & Aug. 16, 2020 &	40	&	40      & HR7596 \\
NGC 7674		&	F23254+0830         & Aug 16-18, 2020 & 40  &   40      & HR7596 \\
NGC 6921		&	20264+2533\_W       & Aug. 17, 18, 2020 & 80	&	80	& HD15318 (=HR718) \\
\hline
\end{tabular}
}
}
\label{log}
\end{table*}

\subsection{Observations and data reduction}

\begin{figure}
\begin{center}
\includegraphics[width=0.4\textwidth]{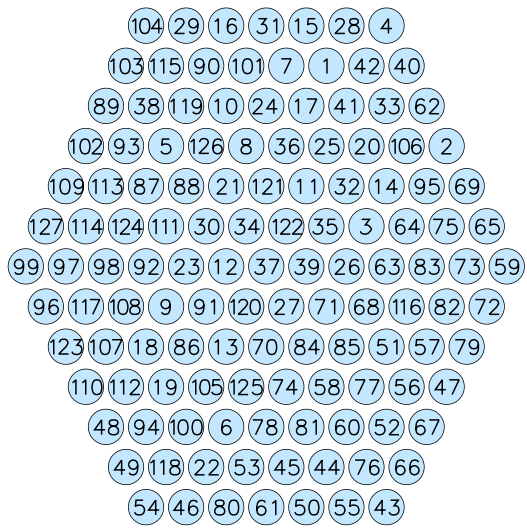}
\end{center}
\caption{The schematic view of KOOLS-IFU with fiber ID.}
\label{fiberID}
\end{figure}

The targets were observed with the KOOLS-IFU on the Seimei Telescope in the 2019B, 2020A, and 2020B semesters (PI: Y.Toba with proposal IDs = 19B-N-CN01, 19B-K-0001, 20A-N-CN01, and 20B-K-0004).
The Seimei is a new 3.8-meter diameter optical-IR alt-azimuth mount telescope located at Okayama Observatory, Kyoto University, Okayama prefecture in Japan, where the typical seeing on this site is 1.2\arcsec--1.4\arcsec.
The science operation was started in 2019.
The KOOLS-IFU consists of 127 fibers with a total field of view (FoV) of $\sim 15\arcsec$ in diameter\footnote{This configuration is valid until September 2020.}.
Each fiber is assigned a unique fiber ID as shown in Figure \ref{fiberID}.
The spatial sampling is $\sim 1.2\arcsec$ per fiber, and the 10$\sigma$ limiting magnitude is 17--18 AB mag given 10 minutes of exposure.
We used the VPH 495 and VPH 683\footnote{We did not use an order-sorting filter O56 that is adopted for blocking the light with $\lambda < 5600$ \AA.} grisms among four grisms equipped with the KOOLS-IFU. 
The wavelength coverage and the spectral resolution ($R = \lambda/\Delta \lambda$) of VPH 495 and VPH 683 are 4300--5900~\AA~and $R$ $\sim$1500 and 5800--8000~\AA ~and $R$ $\sim$2000, respectively\footnote{\url{http://www.o.kwasan.kyoto-u.ac.jp/inst/p-kools/performance/}}.
The observational log is summarized in Table \ref{log}.

The data reduction was executed with a dedicated software\footnote{\url{http://www.o.kwasan.kyoto-u.ac.jp/inst/p-kools/}} in which the Image Reduction and Analysis Facility ({\tt IRAF}: \cite{Tody86,Tody93}) tasks are utilized through the Python package ({\tt PyRAF}: \cite{pyraf}).
In particular, the IRAF {\tt imred.hydra package} \citep{Barden94,Barden95} was mainly used for spectrum extraction, flat fielding, and wavelength calibration in a standard manner.
Following \citet{Matsubayashi}, spectrum extraction was done by measuring the center position of a spectrum at each pixel based on dome flat frames in which we summed the counts within $\pm$2.5 pixels of the center position.
The wavelength was calibrated with arc (Hg and Ne) lamp frames.
The flux calibration was carried out by using an observed spectrum of a standard star (see Table \ref{log}).
Since we utilized all fibers of KOOLS-IFU for targets, we took sky frames separately for sky subtractions.

\subsection{Spectral fitting}
\label{s_specfit}
In order to measure the properties of emission lines such as line flux and line width for our GOALS sample, we conducted a spectral fitting to the reduced spectra by using the Quasar Spectral Fitting package ({\tt QSFit} v1.3.0\footnote{\url{https://qsfit.inaf.it}}; \cite{Calderone}).
Following \citet{Toba_21}, we fit KOOLS-IFU spectra with the following five components; (i) AGN continuum with a single power law, (ii) Balmer continuum modeled by \citet{Grandi} and \citet{Dietrich}, (iii) host galaxy component taken from an empirical SED template in which a Spiral galaxy template\footnote{We note that choice of the galaxy templates does not significantly affect the result of the fitting and our conclusion.} \citep{Polletta} is employed, (iv) iron blended emission lines with UV-optical templates \citep{Vestergaard,Veron}, and (v) emission lines with Gaussian components.
{\tt QSFit} allows us to fit all the components simultaneously based on a Levenberg-Marquardt least-squares minimization algorithm with {\tt MPFIT} \citep{Markwardt} procedure.

Since {\tt QSFit} is optimized to optical spectra taken by the SDSS (e.g., \cite{Calderone,Toba_21,Toba_21b}), we modified the code for general-purpose, which allows us to do the spectral fitting to KOOLS-IFU data.
We fit the Balmer lines with narrow and broad components in which the FWHM of narrow and broad components is constrained in the range of 100--1000 km s$^{-1}$ and 900--15,000 km s$^{-1}$, respectively, to allow a good decomposition of the line profile, in the same manner as \citet{Toba_21}.
Galactic extinction is corrected according to \citet{SF}.
Since one of the purposes of this work is to measure the kinematics of an ionized gas outflow, we also account for a blue-wing component of [O{\,\sc iii}]$\lambda\lambda$4959,5007 lines. 
Following \citet{Calderone}, the FWHM of the broad blue-wing component is constrained to be larger than the narrow component in the range of 100--1000 km s$^{-1}$ while the broad component is allowed to be blue-shifted up to 2000 km s$^{-1}$.
The velocity offsets of [O{\,\sc iii}]$\lambda$4959 and [O{\,\sc iii}]$\lambda$5007 are tied\footnote{This tying could fail when the SN of the emission line is low (see e.g., the middle panels of Figure 4). In this work, we employed the velocity offset of [O{\,\sc iii}]$\lambda$5007 as long as it was reasonably measured without any warning, even if the velocity offset of [O{\,\sc iii}]$\lambda$4959 might not be securely estimated.}.
The measurement errors are estimated from the Monte Carlo resampling method in which we adopted the 1$\sigma$ dispersion of each value by measuring them 100 times for spectra while randomly adding the noise in the same manner as in \citet{Toba_17b} (see Appendix B in \cite{Calderone} for a full explanation of this procedure).

\begin{figure}
\begin{center}
\includegraphics[width=0.48\textwidth]{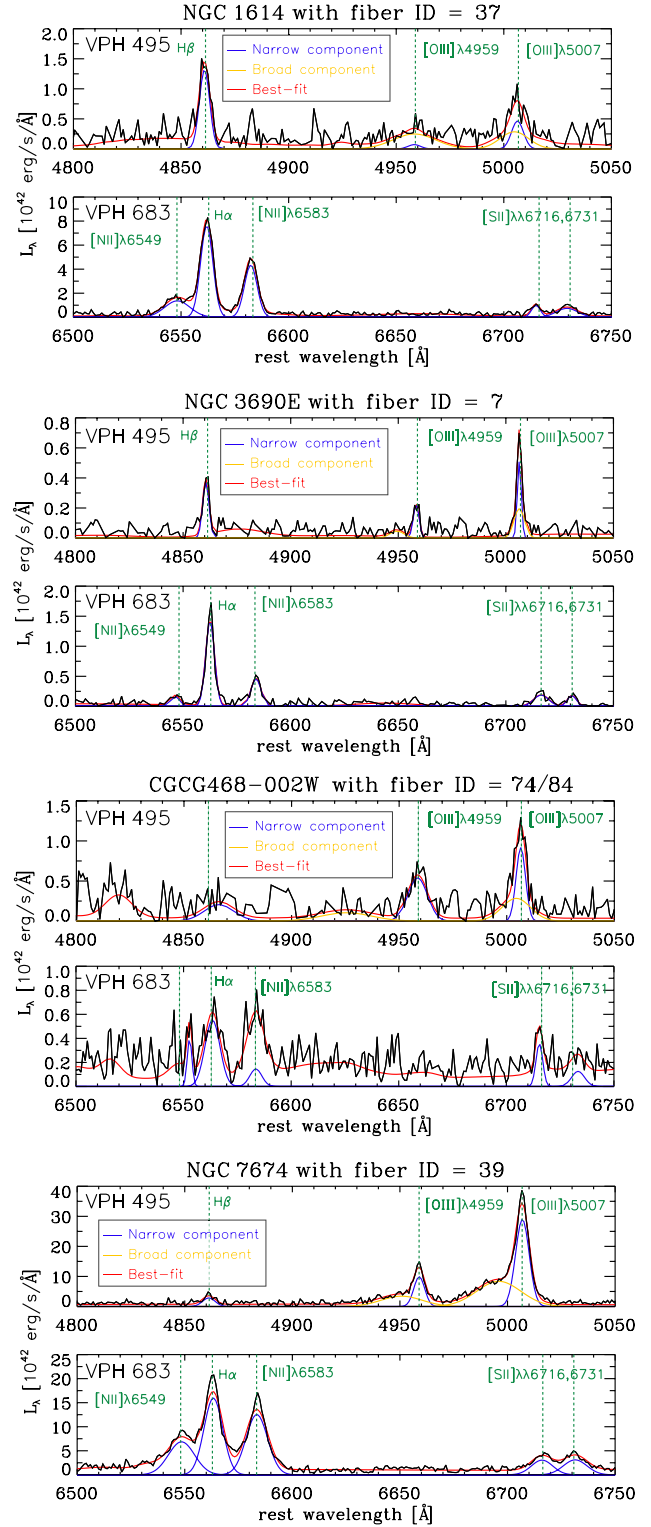}
\end{center}
\caption{Examples of spectral fitting for our objects in fiber with VPH 495 (top) and VPH 683 (bottom). The black lines are observed data. The blue and yellow curves show the narrow and broad components of emission lines, respectively. The best-fit spectrum is shown with red lines. The vertical green dashed lines correspond to the rest-frame wavelengths for H$\beta$, [O{\,\sc iii}]$\lambda\lambda$4959,5007, [N{\,\sc ii}]$\lambda\lambda$6549,6583, H$\alpha$, and [S{\,\sc ii}]$\lambda\lambda$6716,6731 lines.}
\label{QSFIT}
\end{figure}

\section{Results and discussion}
\label{s_RD}
\subsection{Result of the spectral fitting}
Figure \ref{QSFIT} shows examples of the optical spectral fitting with {\tt QSFit}. 
We confirm that about 72\% of the fibers are successfully fitted, of which $\sim90$\% have reduced $\chi^2 < 1.2$. 
This indicates that the spectra of our sample taken by KOOLS-IFU are well fitted by {\tt QSFit}.
The failure of the fitting is mainly due to the low SN of spectra.

\subsection{Line luminosity maps and line ratio diagnostics}
\label{s_BPT}

\begin{figure*}
\begin{center}
\includegraphics[width=0.85\textwidth]{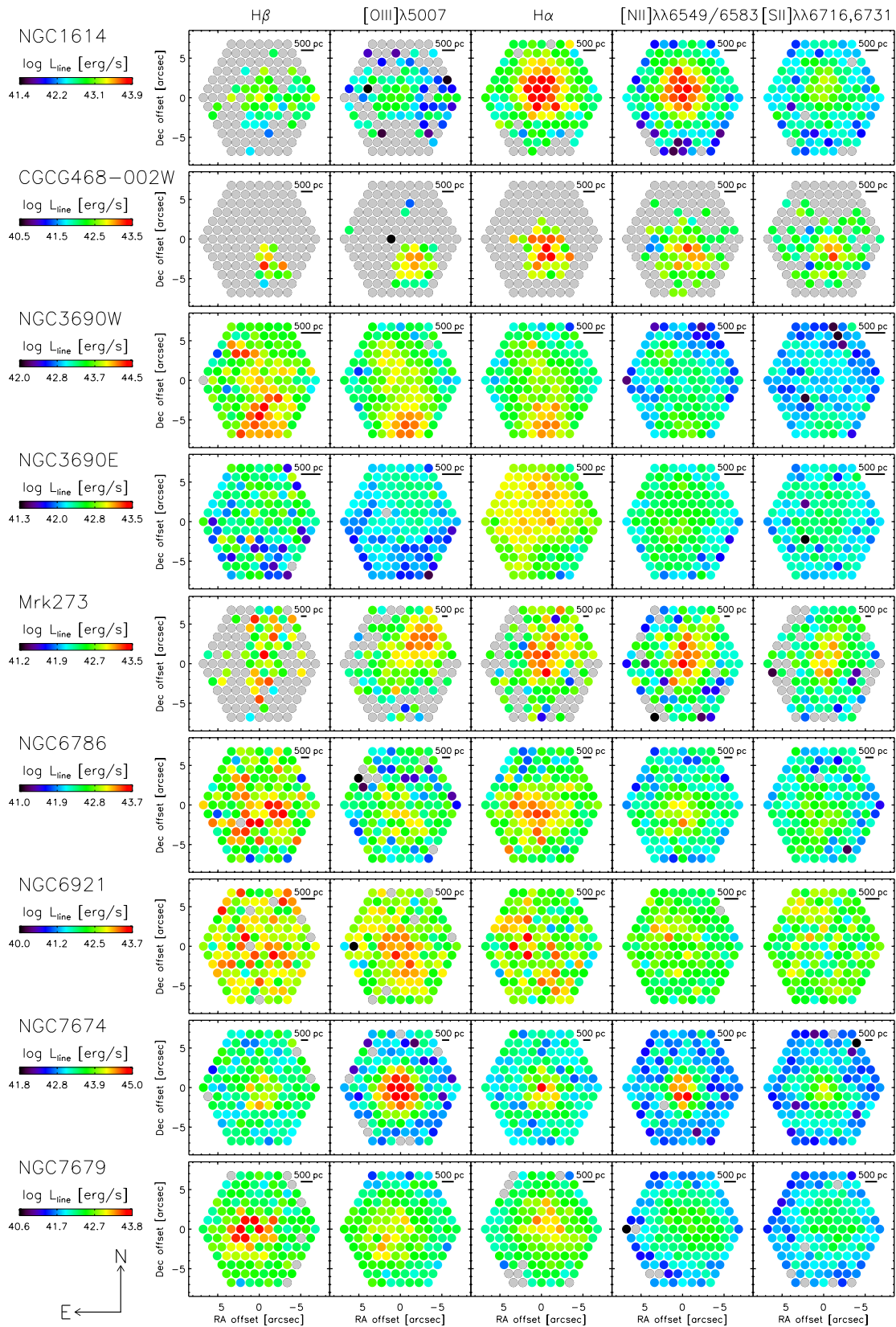}
\end{center}
\caption{Maps for several line luminosities (H$\beta$, [O{\,\sc iii}]$\lambda$5007, H$\alpha$, [N{\,\sc ii}]$\lambda\lambda$6549,6583, and [S{\,\sc ii}]$\lambda\lambda$6716,6731 from left to right) for our sample. Gray circles denote fibers for which line luminosity could not be estimated due to poor SN. North is up and east is left in all images.} 
\label{intensity}
\end{figure*}

Figure \ref{intensity} shows maps for line luminosity in units of erg s$^{-1}$.
Luminosity is the sum of the broad and narrow emission line components as long as a broad component is detected.
Otherwise, a narrow component of the emission line is used for estimating the line luminosity.
Since each spectrum in fiber is fitted 100 times via the Monte Carlo realization (see Section \ref{s_specfit}), their weighted mean is used. 
Line luminosity is successfully estimated for over 95\% of fibers for most objects.
We note that we did not correct for internal dust extinction because it is quite hard to estimate the amount of attenuation through the Balmer decrement in all fibers due to poor SN of H$\alpha$ and/or H$\beta$ line in some fibers (see Figure \ref{intensity}).
But this does not affect estimating the outflow properties that are relevant to the conclusion of this work (see Section \ref{D_dist}).

\begin{figure}
\begin{center}
\includegraphics[width=0.45\textwidth]{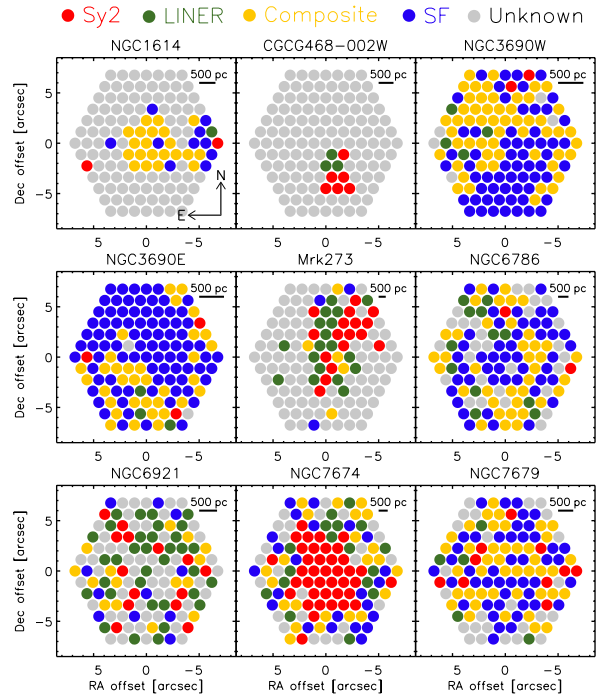}
\end{center}
\caption{Spatially resolved BPT diagram for our sample. Red, green, yellow, and blue circles correspond to Sy2, LINER, composite, and SF, respectively. Gray circles denote ``Unknown'' fibers for which line ratio diagnostics could not be executed because either or both lines for line ratio were undetected. North is up and east is left in all images.}
\label{BPT}
\end{figure}

Figure \ref{BPT} shows a spatially resolved BPT diagram \citep{BPT} in which each fiber position is classified as Seyfert 2 (Sy2), low-ionization nuclear emission-line region (LINER), composite, star-forming region (SF), and Unknown.
We employed diagnostics with line ratios of [N{\,\sc ii}]$\lambda$6583/H$\alpha$ and [O{\,\sc iii}]$\lambda$5007/H$\beta$ provided by \citet{Kauffmann} and \citet{Kewley} (see also \cite{Toba_13,Toba_14}), in which narrow component of each emission line is used.
If we define a fiber with Sy2 or composite classification as AGN-dominant, NGC 1614, CGCG 468-002W, Mrk 273, and NGC 7674 are AGN-dominated with 50-70\% of fibers being classified as Sy2/Composite.
For NGC 3690 West and East (that is classified as merger stage C), about 50\% and 30\% of fibers are classified as AGN, respectively, which suggests that NGC 3690 is basically SF-dominated, but the AGN contribution to the West part is larger than that of the East part for NGC 3690.
Indeed, \citet{Alonso00} reported an interaction-induced star formation for this system.
Hard X-ray observations also supported the presence of AGN in NGC 3690 West and the absence of AGN (or very faint AGN) in NGC 3690 East \citep{Yamada21}.
A caution here is that obscured AGN often lacks the signature of AGN emission from narrow-line regions (NLRs) in optical spectra because NLR is not well developed, and hence those are even outside of the AGN region in the BPT diagram (\cite{Hickox}, and references therein), which makes the diagnostics with line ratios difficult.

\subsection{Outflow properties as a function of distance from the galaxy center}
\label{D_dist}
We examine the spatial correlation of outflow properties in a galaxy and how the outflow strength could be associated with the merger stage.
The velocity offset and/or velocity dispersion of [O{\,\sc iii}]$\lambda$5007 have been preferentially used for evaluating the strength of ionized gas outflow from the rest-frame optical spectra (e.g., \cite{Zakamska,Rakshit,Chen,Jun}).
But it is known that the velocity offset or velocity dispersion alone is not always a good tracer of the strength of AGN outflows because they are often affected by dust extinction \citep{Bae14,Woo}.
To make a fair comparison of outflow power for objects in any merger stage that may have different amounts of dust (e.g., \cite{Ricci,Blecha,Ricci21,Yamada21}), we employ the following quantity;
\begin{equation}
\sigma_0 = \sqrt{v_{\rm [OIII]}^2 + \sigma_{\rm [OIII]}^2},
\end{equation}
where $v_{\rm [OIII]}$ and $\sigma_{\rm [OIII]}$ are velocity shifts of the broad blue-wing component and its velocity dispersion, respectively.
This quantity minimizes the influence of dust extinction; \citet{Bae16} demonstrated that $\sigma_0$ does not change significantly regardless of the amount of dust extinction given a bicone inclination by assuming a biconical outflow model (see \cite{Bae16} for details).
This methodology is applicable even to heavily dust-obscured galaxies \citep{Toba_17b}.

\begin{figure}
\begin{center}
\includegraphics[width=0.45\textwidth]{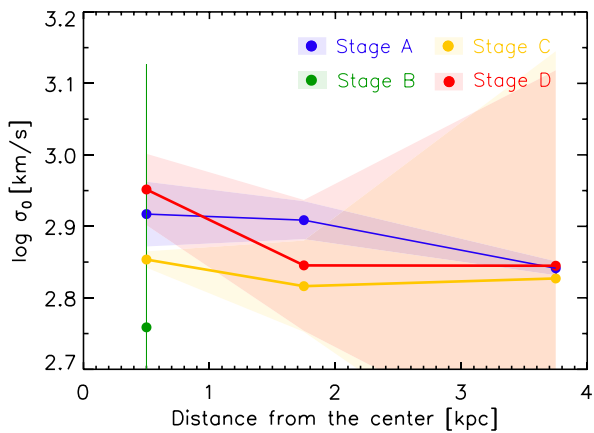}
\end{center}
\caption{The outflow strength ($\sigma_0$) as a function of the galaxy (fiber) center. Blue, green, yellow, and red data points represent the merger stage of A, B, C, and D, respectively.}
\label{sigma_0_dist}
\end{figure}

We then investigate how $\sigma_0$ could depend on the distance from the galaxy center (i.e., fiber center).
In order to obtain reliable $\sigma_0$ with a relatively small uncertainty, we estimate $\sigma_0$ in the following three annular areas with an inner-outer radius of (i) 0--1 kpc, (ii) 1--2 kpc, and (iii) 2--5 kpc. 
Figure \ref{sigma_0_dist} shows resultant $\sigma_0$ as a function of galaxy center for various merger stages.
The only object classified as merger stage ``B'',  CGCG 468$-$002W, is shown only in the innermost area because $\sigma_0$ could not be measured in the outer regions due to poor SN\footnote{Even in the innermost area, the uncertainty of $\sigma_0$ is about 0.3 dex (see Figure \ref{sigma_0_dist}).}.
We find a negative correlation between distance from the center and $\sigma_0$ regardless of the merger stage, with a coefficient of $r \sim -0.4$ in which uncertainties of $\sigma_0$ are taken into account (see e.g., \cite{Kelly,Toba_19}).
This result supports an AGN-driven outflow; an ionized gas outflow is launched from the galactic nucleus.
Focusing on the galactic center, $\sigma_0$ in merger stage D is the strongest that is comparable to dust-obscured AGN \citep{Toba_17b} (see also Section \ref{D_stage}).

It should be noted that a starburst-driven outflow is also reported in some galaxies and is reproduced by numerical simulations (e.g., \cite{Jogee,Leon,Schneider}).
Hence $\sigma_0$ may not be tracing purely AGN-driven outflow.
To test this possibility, we utilize SDSS spectra of nearby starburst galaxies and execute the spectral analysis.
We first select 430 objects from the SDSS DR17 with satisfying the following criteria; (i) $0.01 < z < 0.04$, (ii) $r$-mag $< 15$, and (iii) {\tt SUBCLASS} = ``STARBURST''\footnote{It labels a galaxy with being classified  star-forming galaxies (SFGs) in the BPT diagram but has an equivalent width of H$\alpha$ greater than 50 \AA.}.
The redshift and magnitude ranges are similar to those in our GOALS sample.
The fiber diameter of the SDSS/BOSS Spectrograph is 2\arcsec--3\arcsec, which corresponds to central 2--3 fibers in KOOLS-IFU.
We then perform the spectral fitting with {\tt QSFit} in the same manner as what is described in Section \ref{s_specfit} and estimate $\sigma_0$.
As a result, a weighted mean of $\sigma_0$ for the starburst galaxy sample is $\log\,\sigma_0 \sim 2.7$.
This value is smaller than those in the inner part of our GOALS sample, which means that it could be difficult for starburst galaxies to produce such a strong outflow with $\log\,\sigma_0 > 2.8$ as we observed for our sample (Figure \ref{sigma_0_dist}).
This suggests that high $\sigma_0$ basically originates from an AGN-driven outflow.

In order to see how the galaxy merger could affect the gas density from the spatially-resolved view, we also estimate electron density ($n_{\rm e}$) of ionized gas from the line flux ratio of [S{\,\sc ii}]$\lambda\lambda$6716,6731 that is known as a good tracer of $n_{\rm e}$ and widely used for AGN\footnote{Recently, some caveats of using [S{\,\sc ii}] doublet for $n_{\rm e}$ have been reported (\cite{Davies}, and references therein).} (e.g., \cite{Osterbrock,Kawasaki,Kawaguchi,Joh}), which tells us how $n_{\rm e}$ could be different among the inner and outer regions of our GOALS sample. 
Here, a conversion formula is employed to estimate $n_{\rm e}$ from the [S{\,\sc ii}] ratio (see e.g., \cite{Kakkad}, and reference therein).
Albeit with a large uncertainty ($\sim$30 \%), we find a spatial gradient of $n_{\rm e}$, i.e., high-density gas with $n_{\rm e} \sim 10^3$ cm$^{-3}$ is located in the central 0.5 kpc while low-density gas with $n_{\rm e} < 100$ cm$^{-3}$ is distributed in the outer region.

\subsection{Outflow properties as a function of merger stage}
\label{D_stage}
\begin{figure}[h]
\begin{center}
\includegraphics[width=0.45\textwidth]{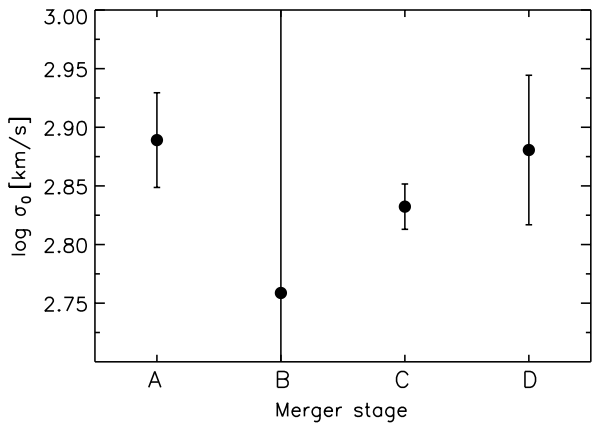}
\end{center}
\caption{The outflow strength ($\sigma_0$) as a function of the merger stage. Since especially objects classified as stage A might be affected by the selection effect or even classified as stage D, our conclusions are based primarily on results from stage B to D (see the text for details).}
\label{sigma_0_mstage}
\end{figure}

Finally, we investigate how $\sigma_0$ could be associated with the merger stages as shown in Figure \ref{sigma_0_mstage}.
We find that ionized gas outflow is more powerful as a sequence of merger stages; the outflow strength in the late-stage (stage D) mergers is about 1.5 times stronger than that in the early-state (stage B) mergers.
We also check the correlation between $\sigma_0$ and $D_{\rm 12}$ in which $D_{\rm 12}$ is the separation between the two nuclei in units of kpc (see \cite{Yamada21}).
We confirm a negative correlation between two quantities for samples with merger stages from B to D.
We further find that $n_{\rm e}$ estimated from the [S{\,\sc ii}] doublet (see Section \ref{D_dist}) seems larger in the late-stage merger. 
Recently, \citet{Yamada21} reported that the outflow velocity of molecular gas of X-ray AGN in merger stage D correlates with Eddington ratio\footnote{\citet{Yamada21} also reported a positive correlation between outflow velocity of molecular gas and X-ray bolometric correction that may be correlated with Eddington ratio (see e.g., \cite{Toba_19}).}.
Molecular outflows are also correlated with ionized gas outflow regarding the mass outflow rate \citep{Fiore}.
These results suggest that the galaxy merger could induce a dense ionized gas outflow driven by AGN, particularly in the late-stage merger.
At the same time, an important caveat to be kept in mind is that those results are based only on nine objects. 
We cannot rule out the possibility that the observed trends are affected by a selection effect (see below). 
Further observations for the GOALS sample with a wider range of redshift, brightness, and dynamical state are needed to corroborate our conclusion.

We also note that $\sigma_0$ for merger stage A is comparably large to stage D.
One possible reason for this is that merger stage A is a too early phase (i.e., two galaxies are too far from each other) to affect the strength of the ionized gas outflow, and thus $\sigma_0$ could depend on the nature of the individual object rather than the contribution of the merger process.
Indeed, NGC 6921, NGC 7674, and NGC 7679 (that are classified as merger stage A) are all reported as strong AGNs with X-ray detections\footnote{NGC 7674 and MGC 7679 are also reported [Ne{\,\sc v}]$\lambda$14.32 $\mu$m detections \citep{Inami}.} (see \cite{Yamada21}, for details).
In particular, NGC 7674 is expected to harbor dual SMBHs (\cite{Kharb}, but see also \cite{Breiding} for counterargument), which could have a quite strong AGN outflow\footnote{The Eddington ratio of NGC 7674 is also larger than other AGN in our sample \citep{Yamada21}.}.
A full investigation of this object will be reported (S.Yamada et al. in preparation).
Another possibility is that those objects are indeed at the very last stage of former merging events so that no tidal features are visible (which means they are stage D).
On the other hand, if the trend in Figure \ref{sigma_0_mstage} is robust without being affected by the selection effect, this result could suggest that galaxy merger is not the only path to ignite AGN as introduced in Section \ref{Intro}.

\citet{Arribas_14} reported that SF-driven outflow strength traced by H$\alpha$ line in SFGs is correlated with dynamical phases (that corresponds to the merger stage, see \cite{Rodriguez}, for more detail) along the merging process.
\citet{Rich} also reported from an IFU observation that the low-velocity shocks become increasingly prominent as a merger progresses 
(see also \cite{Rich11}).
In addition, the outflow strength ($\sigma_0$) seems to be correlated with the Eddington ratio (e.g., \cite{Bae14,Toba_17b}).
Taking account of those findings into account, our observations support that galaxy mergers enhance both SF and AGN activity and play an important role in the co-evolution of galaxies and SMBHs.

\section{Summary}
\label{s_C}
In order to reveal how the galaxy merger could affect the strength of the ionized gas outflow from a spatially-resolved point of view, we analyze 9 local U/LIRGs at $z < 0.04$ through the IFU observations with KOOLS-IFU on Seimei Telescope.
The observed targets are selected from a 60 $\mu$m flux-limited sample in the GOALS project, covering a wide range of merger states from an early-stage (A) to late-stage (D).
We successfully detect various emission lines of H$\beta$, [O{\,\sc iii}]$\lambda\lambda$4959,5007, H$\alpha$, [N{\,\sc ii}]$\lambda\lambda$6549,6583, and [S{\,\sc ii}]$\lambda\lambda$6716,6731 over 70\% of fibers.
This work mainly focuses on the strength of the ionized gas outflow ($\sigma_0$) by using a combination of velocity shift and dispersion of [O{\,\sc iii}]$\lambda$5007 line.
With all the caveats discussed in Sections \ref{D_dist} and \ref{D_stage} in mind, we find $\sigma_0$ shows a negative correlation with the distance from the galaxy center and a positive correlation with a sequence of merger stages. 
This indicates that galaxy mergers may induce AGN-driven outflow and maximize the late-stage merger, which is in good agreement with previous works reporting that the Eddington ratio of AGN increases along with the merger stage. 

Since our follow-up observation campaign with KOOLS-IFU is ongoing, and about 130 GOALS samples are observable from the Seimei telescope, this work provides a benchmark for revealing the spatially-resolved outflow properties from a statical point of view in the future. 

\begin{ack}
We gratefully thank the anonymous referee for a careful reading of the manuscript and very helpful comments.
We acknowledge Dr. Kyuseok Oh for providing a finding chart for our observations in 2019.
This work is supported by JSPS KAKENHI Grant numbers 18J01050, 19K14759, and 22H01266 (YT), 19J22216 (SY), 20H01946 (YU), and 20K04027 (NO).
SY is grateful for support from RIKEN Special Postdoctoral Researcher Program.
\end{ack}

\end{document}